\documentclass[a4paper, amsfonts, amssymb, amsmath, reprint, showkeys, nofootinbib, twoside]{revtex4-1}
\usepackage[english]{babel}
\usepackage[utf8]{inputenc}
\usepackage[colorinlistoftodos, color=green!40, prependcaption]{todonotes}
\usepackage{amsthm}
\usepackage{mathtools}
\usepackage{physics}
\usepackage{xcolor}
\usepackage{graphicx}
\usepackage[left=23mm,right=13mm,top=35mm,columnsep=15pt]{geometry} 
\usepackage{adjustbox}
\usepackage{placeins}
\usepackage[T1]{fontenc}
\usepackage{lipsum}
\usepackage{csquotes}
\usepackage{setspace}
\usepackage{siunitx}
\usepackage[english]{babel}
\usepackage[utf8]{inputenc}
\usepackage[pdftex, pdftitle={Article}, pdfauthor={Author}]{hyperref}

\setcitestyle{super,compress}

\newcommand{\NEW}[1]{\textcolor{black}{#1}}

\begin{document}

\title{Structural Color from Solid-State Polymerization-Induced Phase Separation}

\author{Alba Sicher$^{1,2}$, Rabea Ganz$^{1}$, Andreas Menzel$^{3}$, Daniel Messmer$^{4}$, Guido Panzarasa$^{1}$, Maria Feofilova$^{1}$, Richard O. Prum$^{5}$, \NEW{Robert W. Style$^{1}$}, Vinodkumar Saranathan$^{6}$, Ren\'{e} M. Rossi$^{2*}$, Eric R. Dufresne$^{1*}$}
 \affiliation{\vspace{0.5cm}$^{1}$Laboratory for Soft and Living Materials, Department of Materials, ETH Z\"{u}rich, 8093 Z\"{u}rich, Switzerland.}
 \affiliation{$^{2}$Laboratory for Biomimetic Membranes and Textiles, Empa, Swiss Federal Laboratories for Materials Science and Technology, 9014 St. Gallen, Switzerland.}
 \affiliation{$^{3}$Paul Scherrer Institut, Forschungsstrasse 111, 5232 Villigen PSI, Switzerland.}
 \affiliation{$^{4}$Laboratory of Polymeric Materials, Department of Materials, ETH Z\"{u}rich, 8093 Z\"{u}rich, Switzerland.}
 \affiliation{$^{5}$Department of Ecology and Evolutionary Biology and the Peabody Museum, Yale University, New Haven, CT 06520.}
 \affiliation{$^{6}$Division of Science, Yale-NUS College, 10 College Avenue West, 138609, Singapore.}
 \affiliation{}

 \email[Email: ]{rene.rossi@empa.ch, eric.dufresne@mat.ethz.ch}

\date{\today} 
\vspace{1.6cm}

\begin{abstract}
\vspace{0.2cm} Structural colors are produced by wavelength-dependent scattering of light from nanostructures.
While living organisms often exploit phase separation to directly assemble structurally colored materials from macromolecules,
synthetic structural colors are typically produced in a two-step process involving the sequential synthesis and assembly of building blocks. 
Phase separation is attractive for its simplicity, but applications are limited due to a lack of robust methods for its control.
A central challenge is to arrest phase separation at the desired length scale.
Here, we show that solid-state polymerization-induced phase separation can produce stable structures at optical length scales. 
In this process, a polymeric solid is swollen and softened with a second monomer.
During its polymerization, the two polymers become immiscible and phase separate.
As free monomer is depleted, the host matrix resolidifies and arrests coarsening.
The resulting polymeric composites have a  blue or white appearance.
We compare these biomimetic nanostructures to those in structurally-colored feather barbs, and demonstrate the flexibility of this approach by producing structural color in filaments and large sheets.
\end{abstract}


\maketitle

\par 
In conventional dyes and pigments,  color is encoded in the electronic structure of molecules. 
In structurally colored materials, color emerges from the interference of light scattered from sub-micron variations in the refractive index \cite{{parker2000515},{parker2006structural},{rayleigh1919vii},{noh2010noniridescent},{liew2011short}}.
Since a wide spectrum of structural colors can be made by different nanostructures with the same chemical composition \cite{{saranathan2012structure},{zi2003coloration},{teyssier2015photonic}}, structural color has the potential for more sustainable color production.

\par Colloidal processing and block-copolymer assembly have emerged as leading methods to produce synthetic  structurally colored materials \cite{{nagayama1996two},{jiang1999single},{matsubara2007thermally},{forster2010biomimetic},{finlayson2011ordering},{ge2014angle},{xiang2016new},{lee2017chameleon},{fu2017bio},{han2019cascade},{shang2020photonic},{edrington2001polymer},{valkama2004self},{boyle2017structural},{appold2018one},{vatankhah2018chameleon},{wu2018flexible},{patel2020tunable}}. 
In all these cases, the characteristic length scale of the final structure is determined by the size of its  building block.  
The fabrication process involves  two distinct steps: the synthesis of  building blocks and their subsequent assembly.  
Synthesis and assembly typically have contradictory physical and chemical parameters, necessitating additional processing and purification steps between them.
For example, synthesis requires a low volume fraction of particles, while assembly requires a high one.

\par In contrast, living systems construct photonic structures directly from macromolecules, whose dimensions are much smaller than the characteristic scale of the final structure.
In structurally-colored bird feather barbs, phase separation is triggered by the supra-molecular polymerization of  
 cytoplasmic $\beta$-keratin into  filaments  \cite{prum2009development,dufresne2009self}.
Phase separation is ultimately arrested by  intermolecular cross-linking \NEW{of} disulfide bonds, leaving stable nanostructures composed of air and cross-linked protein. 
 These  structures come in two characteristic morphologies:  dense packings of monodisperse air bubbles or  \NEW{bi-continuous channels} \cite{{saranathan2012structure}}. 
These composites have pronounced short-range order,  and reflect a narrow band of wavelengths \cite{noh2010noniridescent,prum2013fourier,prum2003coherent,prum2003structural}.

\par Recently, scientists have begun to exploit phase separation to spontaneously form photonic structures. 
Strongly white networks have been obtained from quenching polymer-solvent mixtures \cite{syurik2018bio}, and thin structurally-colored films have been produced in an immiscible polymer blend through a 2D phase-separation process triggered by the evaporation of a co-solvent \cite{nallapaneni2017specular}.
Structural colors have also been observed in gels of a single polymer,  formed by polymerization-induced phase separation   \cite{kumano2011tunable}.

\par Here, we produce structural colors by polymerization-induced phase separation (PIPS) in the solid state.
Starting with a solid block of uncrosslinked polystrene (PS) swollen with methyl methacrylate monomer (MMA), we initiate free-radical polymerization and subsequent phase separation. 
As monomer is depleted from the system, the matrix re-solidifies, arresting the coarsening process. 
This biologically-inspired process produces inclusions with a modest degree of short-range order.
The resulting materials have structural blue to white appearances.
We demonstrate the potential scalability and versatility of solid-state PIPS by making large structurally colored films and filaments.

\begin{figure*}[h!tb] 
\centering 
\includegraphics[width=17cm]{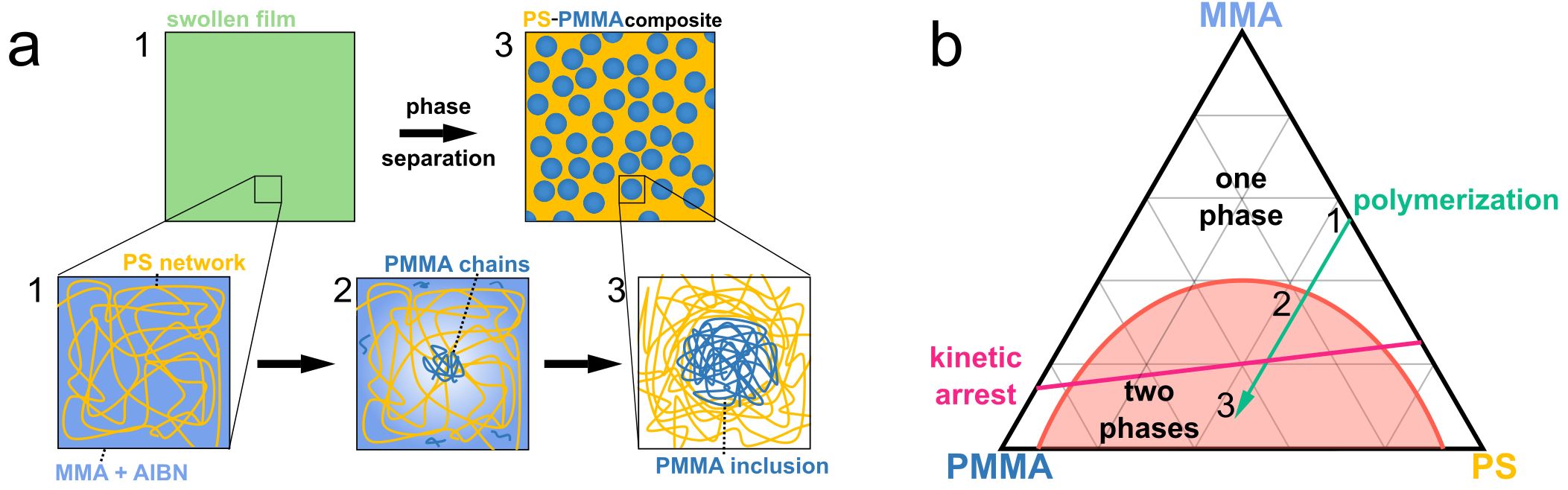}
\caption{\emph{Solid-state polymerization-induced phase separation}. \NEW{a) Schematic representation the process. Microscopic (top) and macromolecular (bottom) views, showing the system after swelling (1), at the onset of phase separation (2), and near completion (3). PS (yellow) is swollen with a mixture (blue) of monomer (MMA) and thermal initiator (AIBN).  b) Qualitative phase diagram.
The red line indicates the phase boundary (binodal) and the beginning of a two-phases region. The pink line indicates the onset of kinetic arrest. The typical polymerization procedure is described by the green arrow. Steps 1, 2, and 3 from a) are marked on the diagram.}
}
\label{fig:schematic}
\end{figure*}

\par Our process is schematized in Figure \ref{fig:schematic}.
First, we swell transparent polystyrene (PS)  with methyl methacrylate (MMA) and a thermal initiator (2,2'-Azobis(2-methylpropionitrile) (AIBN))  (Figure \ref{fig:schematic}a,b). 
To fix the quantity of monomer and initiator,  PS films equilibrate overnight in a solution of ethanol, MMA and AIBN. 
 While uncrosslinked PS completely dissolves in a solution of MMA and AIBN, ethanol is a bad solvent for PS.
 The final concentration of MMA in the PS is tuned from 1 to 40 \%wt changing the concentration of MMA in ethanol from 12 to 41 \%wt.  
 \NEW{This corresponds to step 1, indicated in Fig. \ref{fig:schematic}}.
 Then,  we initiate polymerization by placing the samples, still in their bath of swelling fluid, into an oven at a target polymerization temperature, $T_p$, from 60 to 80 $^\circ$C.
As poly(methyl methacrylate) (PMMA) chains grow, their solubility is reduced, favoring phase separation, \NEW{indicated as step 2 in Fig. \ref{fig:schematic}}.
As polymerization proceeds, monomer is depleted and the matrix stiffens. 
\NEW{Eventually,  polymerization goes to near completion, as indicated by step 3 in Fig. \ref{fig:schematic}.
In this final state, the kinetically arrested} matrix suppresses coarsening, and produces stable far-from equilibrium micro-phase-separated structures.

\begin{figure}[bt] 
\centering
\includegraphics[width=8cm]{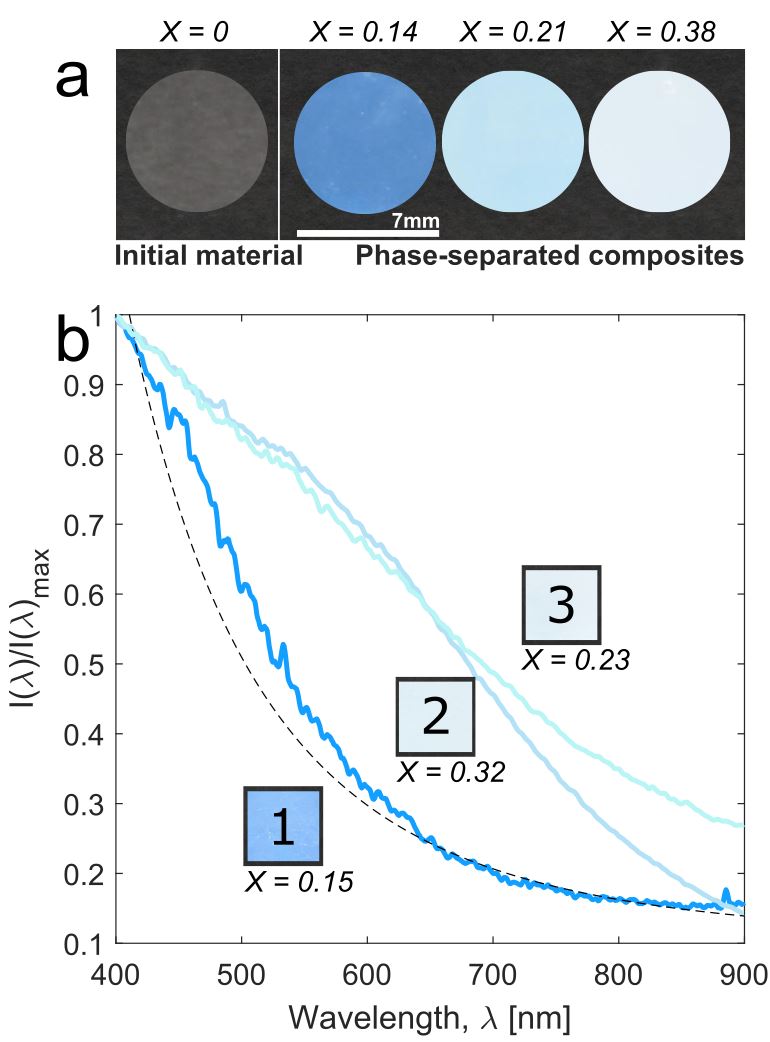}
\caption{\emph{Solid-state polymerization-induced phase separation produces structural blue and white.} a) Photographs of initially transparent PS ($X=0$) and three samples with increasing PMMA fraction $(X=0.14,~0.21,~0.38)$ after polymerization-induced phase separation at 70 $^\circ$C.  b) Intensities as a function of wavelength, $I(\lambda)$, from reflectivity measurements of blue, light blue and white samples. Insets: photographs of the corresponding samples on a black background. The PMMA fraction of each sample is reported. Dashed line: $I\propto\lambda^{-4}$, a Rayleigh scattering fit to long wavelengths of sample 1.
}
\label{fig:phi}
\end{figure}

\par These micro-phase separated composites appear either blue or white depending on the synthesis parameters used. 
Generally, as the fraction of PMMA increases in the sample, its appearance evolves from transparent, to blue, to white (Figure \ref{fig:phi}a).
We quantify the molar fraction of PMMA in the samples with $^{1}$H nuclear magnetic resonance (NMR) spectroscopy, and we define it as 
$X=c_\mathrm{PMMA}/(c_\mathrm{PMMA}+c_\mathrm{PS})$, where $c_\mathrm{PMMA}$ and $c_\mathrm{PS}$ are  the \NEW{molar} concentrations  of polymerized MMA and styrene, respectively.
The final mole fraction of polymerized monomer,
$X$, depends on the concentration of MMA used, as well as the polymerization temperature, and ranges from $\approx 0.1$ to $\approx 0.4$ in these experiments (Table \ref{tab:parameters}).

\begin{table*}[hbt]
\centering
    \begin{tabular}{l|c|c|c|c|c|r}
    \hline
    \textbf{Sample name} & \textbf{\emph{X}} & \textbf{\emph{$\eta$}} & \textbf{$\mathrm{\%_v}$ MMA} & \textbf{$T_p$ [$^\circ$C]} & \textbf{$t_p$ [min]} & \textbf{Figure} \\
    \hline
    Blue & 0.14 & 0.85 & 30 & 70 & 90 & \ref{fig:phi}a \\
    Light Blue & 0.21 & 0.91 & 35 & 70 & 90 & \ref{fig:phi}a \\
    White & 0.38 & 0.96 & 45 & 70 & 90 & \ref{fig:phi}a \\
    \#1 & 0.15 & 0.67 & 40 & 60 & 90 & \ref{fig:phi}b, \ref{fig:stemsaxs}a-c \\
    \#2 & 0.32 & 0.82 & 45 & 60 & 90 & \ref{fig:phi}b, \ref{fig:stemsaxs}a-c, \ref{fig:bird}b\\
    \#3 & 0.23 & 0.97 & 40 & 80 & 90 & \ref{fig:phi}b, \ref{fig:stemsaxs}a-c\\
    \#4 & 0.20 & 0.89 & 40 & 60 & 90 & \ref{fig:surface}a-e, \ref{fig:samples}a,c\\
    Black PS & / & / & 30 & 60 & 90 & \ref{fig:samples}b\\
    Fiber & / & / & 35 & 60 & 90 & \ref{fig:samples}d 
    \end{tabular} 
    \caption{List of all the samples, their compositions and their synthesis parameters: \emph{X} = molar fraction of PMMA in the sample, $\eta$ = monomer conversion, $\mathrm{\%_v}$ MMA = percentage of methyl methacrylate in solution, $T_p$ = polymerization temperature, $t_p$ = polymerization time, PS = polystyrene. \vspace{0.2cm}}
    \label{tab:parameters}
\end{table*}

\par To quantify the appearance of the samples, we measure the diffuse reflectivity of 1 mm thick samples using an angle-resolved spectrophotometer.  
We measure the spectra as close as possible to the backscattering direction, at an angle $\theta=\ang{15}$ between the direction of illumination and detection. 
In Figure \ref{fig:phi}b, we compare the normalized intensities as a function of wavelength, $I(\lambda)$, for samples with different molar fractions of PMMA.
In all cases, the diffuse reflectivity of the samples decreases monotonically with wavelength across the visible spectrum.
This qualitatively resembles Rayleigh scattering. 
Indeed, the longest wavelengths at the lowest polymer fractions, $X\leq 0.20$, are reasonably well-fit by the expected $\lambda^{-4}$ scaling, as shown by the dashed line in Figure \ref{fig:phi}b.
However, all samples show an excess of reflectivity beyond this Rayleigh background. 
\NEW{As the PMMA-fraction increases,} the reflectivity increases \NEW{and} the blue colors fade into white.
For samples polymerized at higher temperatures (Sample 3, $X=0.23$), white color appears at lower PMMA-fractions, as shown in Figure \ref{fig:phi} and in Figure S1. 
Figure S2 shows that the color is angle-independent.

\begin{figure*}[hbt] 
\centering 
\includegraphics[width=17cm]{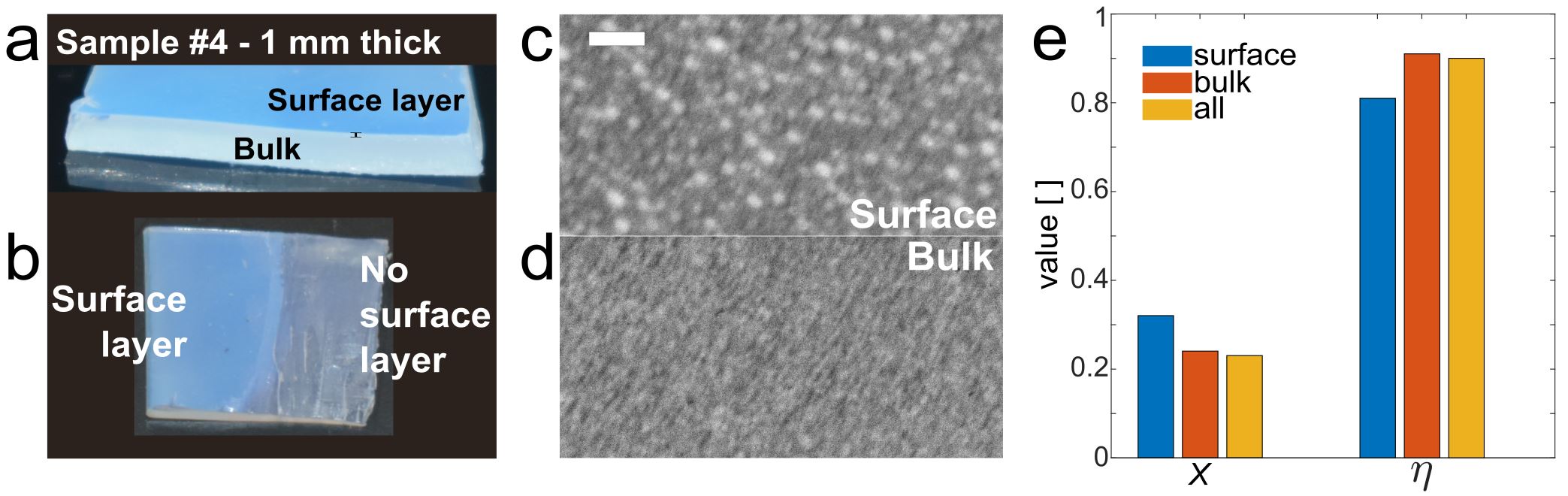}
\caption{\emph{Color originates in a surface layer.}  a) Photograph of a blue sample (Sample 4, $X=0.20$) in cross-section.  b) Photograph of the same sample with the surface layer removed on the right side. c,d) STEM bright-field images of the sample near the surface and in the bulk. Scale bar:  500 nm. e) PMMA fraction and monomer conversion of the sample measured by NMR at the surface, bulk and overall.  
}
\label{fig:surface}
\end{figure*}

\par Color originates from scattering within \NEW{about} $150~\mathrm{\mu m}$ of the surface.
We can see this macroscopically by looking  at the cross-section of a sample  (Figure \ref{fig:surface}a), or by removing the surface layer from the rest of the sample by mechanical abrasion (Figure \ref{fig:surface}b). 
Electron microscopy reveals that the structure near  the surface is distinct from the structure in the bulk.   
Figure \ref{fig:surface}c and d show scanning transmission electron micrographs (STEMs) with the same field of view $4 \times 2~\mathrm{\mu m}^2$ acquired 25 and $270~\mathrm{\mu m}$ from the surface of the same $X=0.20$ sample. 
Near the surface, we find a solid dispersion of PMMA spheres, about hundred nanometers in diameter. In the bulk, we do not observe any phase-separated domains at this length-scale.

\par We quantify the compositional difference between the surface and bulk using NMR. 
As shown in Figure \ref{fig:surface}e, there is an increase in the mole fraction of PMMA, $X$, at the surface relative to the  bulk of the sample. 
We also define the monomer conversion as $\eta=c_\mathrm{PMMA}/(c_\mathrm{MMA}+c_\mathrm{PMMA}$), where $c_\mathrm{MMA}$ and $c_\mathrm{PMMA}$ are the molar concentrations of free MMA and MMA in a polymer. 
In Figure \ref{fig:surface}e, we see that the monomer conversion is lower in the surface than in the bulk.
We hypothesize that the structural difference between surface and bulk arises during polymerization. 
As monomer is depleted in the sample, it can be replenished by the surrounding reservoir of monomer. 
The competition of monomer diffusion and  polymerization kinetics defines a boundary layer where monomer is steadily replenished during polymerization.
This simple picture qualitatively captures \NEW{the observed differences between surface and bulk, and particularly} the complementary increase of $X$ and decrease of $\eta$ in the surface layer.

\begin{figure*}[h!bt] 
\centering
\includegraphics[width=17cm]{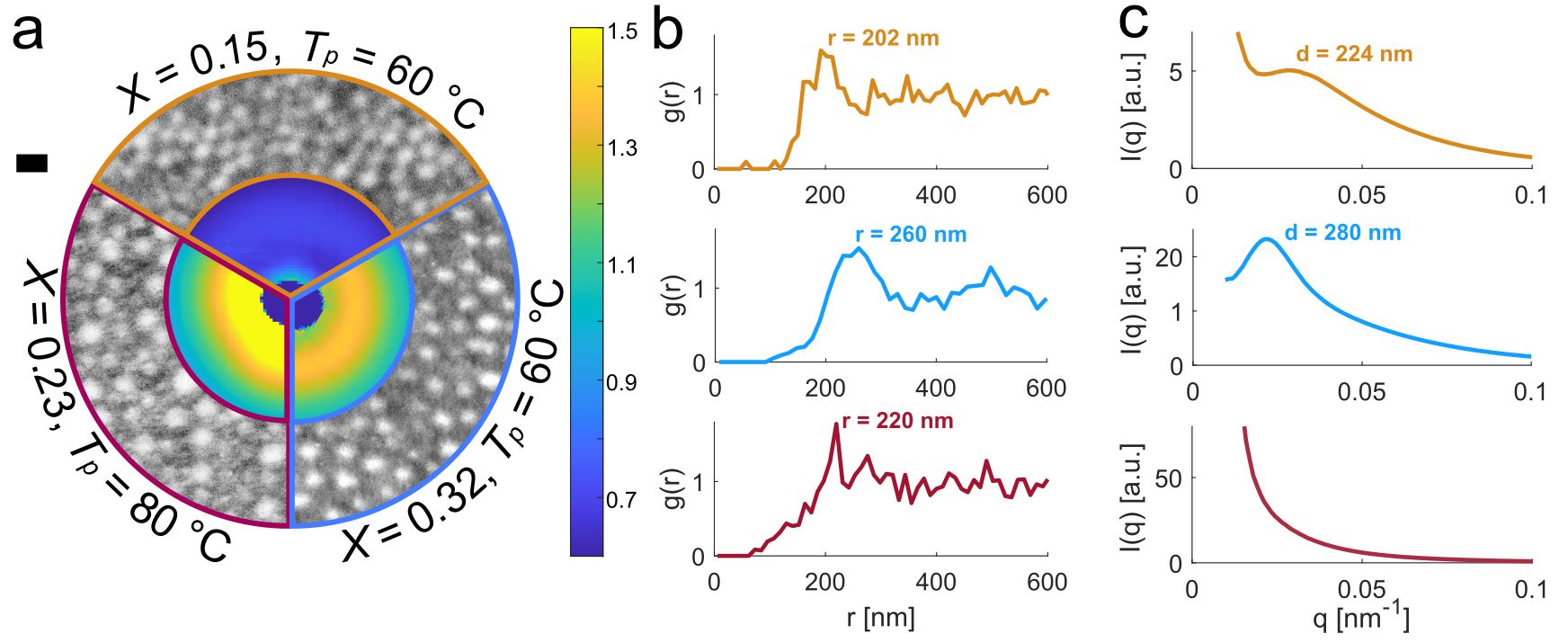}
\caption{\emph{Structural characterization of phase-separated samples.} a) Comparison of three samples with different PMMA fractions, $X$, and polymerization temperatures, $T_p$, as indicated. Outer ring: bright-field STEM images taken in the structured surface layer. Inner ring: sections of SAXS patterns. The dark blue feature in the center is the beamstop. On the right: color bar for the SAXS patterns. 
The scale bar on the left corresponds to 300 nm for the STEM images and $0.009~\mathrm{nm^{-1}}$ for the SAXS patterns. b) Pair correlation functions, $g(r)$, calculated from particle locations extracted from the STEM images in (a). c) Azimuthal averages calculated from the 2D SAXS patterns in (a).}
\label{fig:stemsaxs}
\end{figure*}

\par To determine how the morphology of the surface layer affects the optical properties of the material, we applied both scanning transmission electron microscopy (STEM) and small angle X-ray scattering (SAXS). 
STEM shows how the size, spacing and polydispersity of inclusions depend on the synthesis parameters. 
In the outer ring of Figure \ref{fig:stemsaxs}a, we compare STEM images of three samples with different PMMA fractions and processing temperatures. To quantify the differences in these images,
we locate the centroid of each particle, and calculate the pair-correlation function, $g(r)$, shown in Figure \ref{fig:stemsaxs}b.
All samples show anti-correlation at short spatial scales, presumably due to competition between the PMMA nuclei for free polymer.
Additionally, the 60 $^\circ$C samples show a distinct peak in $g(r)$ at separations of $202~\mathrm{nm}$ and $260~\mathrm{nm}$, with modest peak correlation values of $1.50$ and $1.54$ respectively. 
The other sample, processed at 80 $^\circ$C, shows a hint of a short-ranged positive correlation near the same location, but suffers from limited statistics.
Note that these $g(r)$'s are obtained from 2D sections of a 3D materials, and provide a limited view of the translational order in the samples.

\par Scattering methods precisely quantify 3D structural correlations and provide better statistics by averaging over a much larger sample volume.
The inner ring of Figure \ref{fig:stemsaxs}a shows a sector of these three samples' isotropic SAXS patterns, aligned at $q=0$.
The full scattering patterns can be found in Figure S3.
Azimuthally-averaged scattering profiles are shown in Figure \ref{fig:stemsaxs}c.
While the 60 $^\circ$C samples show modest peaks at $q=0.028~\mathrm{nm^{-1}}$ (Sample 1, $X=0.15$) and $q=0.022~\mathrm{nm^{-1}}$ (Sample 2, $X=0.32$), 
the higher temperature sample shows no significant features across the probed region of the spectrum.
The peak is stronger for Sample 2 ($X=0.32$), whose reflectivity showed a larger deviation from Rayleigh scattering compared to Sample 1 ($X=0.15$) (Figure \ref{fig:phi}).
The peak in the SAXS spectra of the 60 $^\circ$C samples correspond to length scales, \NEW{$d$}, of 224 and \NEW{280 nm, as indicated in Fig. \ref{fig:stemsaxs}c. These values are} similar to the location of the peaks in $g(r)$ calculated from TEM images near the sample surface (Figure \ref{fig:stemsaxs}b).
This suggests that even though the X-ray beam travels across the entire sample, the scattering is dominated by the same near-surface structures.

\par Together SAXS and STEM show weak structural correlations in the samples polymerized at 60 $^\circ$C, dominated by the spacing between the PMMA particles.
Structures formed at higher temperatures consistently show higher polydispersity (Figure S4) and no structural correlations.
This reflects a broadened molecular weight distribution by free-radical polymerization at higher temperatures \cite{{sacks1973effect},{campbell2003high},{walsh2020general}}.

\begin{figure}[bt] 
\centering
\includegraphics[width=8cm]{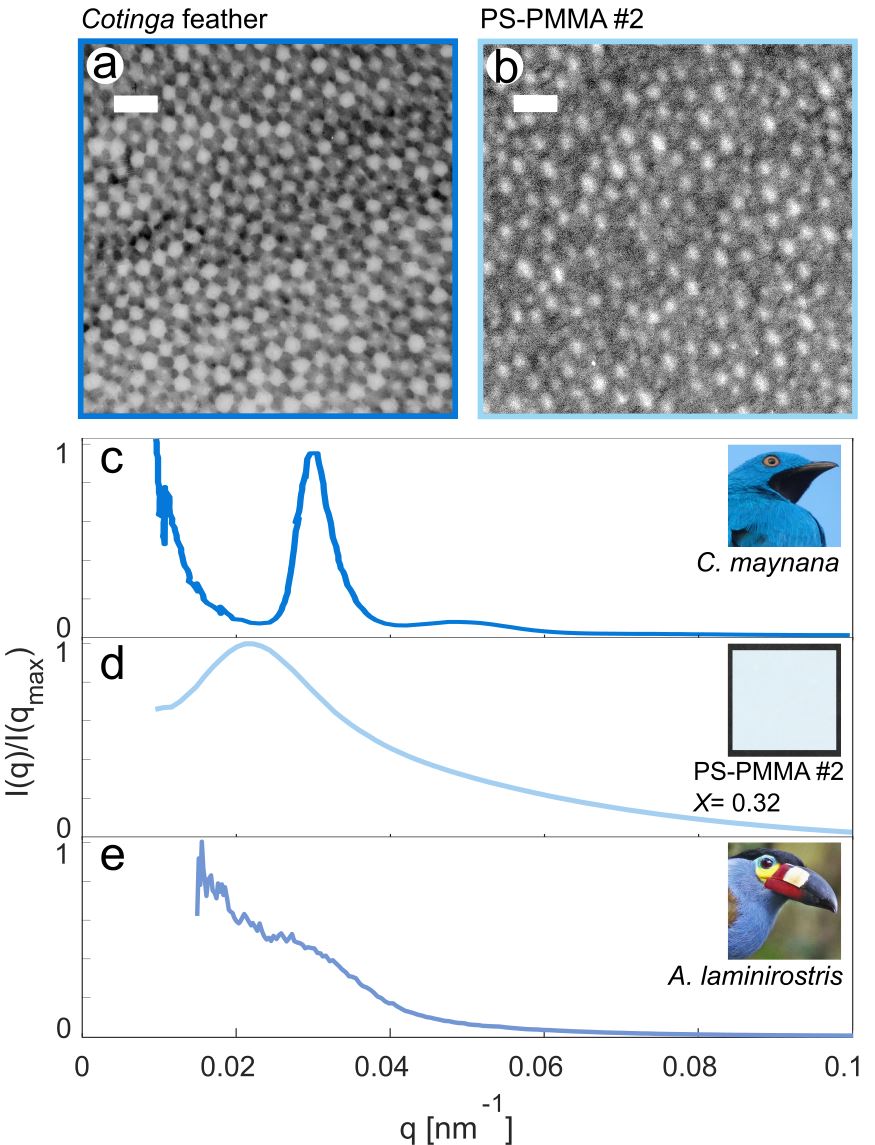}
\caption{\emph{Biological and bio-inspired photonic structures}. a) TEM image of the photonic structure in the feather barbs of \emph{Cotinga maynana}.  b) STEM image of the photonic structure in Sample 2. The images in (a) and (b) have the same field of view. For both the scale bar is 500 nm.  c) Azimuthally-averaged SAXS spectra, $I(q)$, from the feathers of \emph{Cotinga maynana}. Inset: \emph{C. maynana} (image in the public domain). d) $I(q)$ from Sample 2. Inset: photo of Sample 2 on a black background. e) $I(q)$ from the feathers of \emph{Andigena laminirostris}. Inset: \emph{A. laminirostris} (image in the public domain). 
}
\label{fig:bird}
\end{figure}

\par We compare our PS-PMMA structures formed by polymerization-induced phase separation to some of the natural photonic structures which inspired them in Figure \ref{fig:bird}.
Electron micrographs of the color-producing air-keratin structures in the vibrant blue-green feather barbs of \emph{Cotinga maynana} are shown alongside a PS-PMMA sample (Sample 2, $X=0.32$) in Figure \ref{fig:bird}a,b.
Both materials \NEW{have angle-independent blue color, and} show approximately spherical  inclusions.
The inclusions in the feathers, however, are more uniform and more closely packed.
Together, this leads to much stronger structural correlations, as demonstrated by a sharp peak in their SAXS spectrum, shown in Figure \ref{fig:bird}c. 
For comparison, the scattering spectrum of the PS-PMMA sample is shown in Figure \ref{fig:bird}d. 
It features a much broader peak at smaller $q$. 
While the order of our biomimetic samples falls short of the most vibrant structurally colored feather barbs, they are more ordered than others  \cite{saranathan2012structure}.
For example, the SAXS scattering spectrum from the  nanostructures underlying the blue-gray feather barbs of \emph{Andigena laminirostris}
shows only a subtle shoulder on a broad background, as shown in Figure \ref{fig:bird}e.

\par Previous work on structure-property relationships in the color of bird feathers \cite{noh2010noniridescent} has shown that a sharp peak in the SAXS spectrum at $q_\mathrm{max}$ leads to a narrow band in the back-scattering reflection spectrum, centered on $\lambda_\mathrm{max}$,  with $\lambda_\mathrm{max}=4\pi n_e/q_\mathrm{max}$
where $n_e$ is the effective refractive index of the composite.
For bird feathers, $n_e$ is about 1.265 \cite{saranathan2012structure}.  
Thus, the sharp structural correlations of \emph{C. maynana} produce a vivid color at $\lambda_\mathrm{max}\sim 520~\mathrm{nm}$.
The effective index of the PS-PMMA samples can be estimated from the fraction of PMMA as well as the refractive indices of the two phases, $n_\mathrm{PS}=1.57$ and $n_\mathrm{PMMA}=1.48$.
Using the Maxwell-Garnett approximation \cite{{forster2010biomimetic},{garnett1904xii}}, we find effective indices ranging from 1.53 to 1.56.
Applying this analysis, we might expect structural correlations in PS-PMMA Sample 2 to produce some excess scattering over a broad-range of wavelengths in the near infra-red from about 830 to 930 nm.
However, this is overwhelmed by a broad Rayleigh-like background, as shown in Figure \ref{fig:phi}.

\begin{figure*}[bt] 
\centering
\includegraphics[width=17cm]{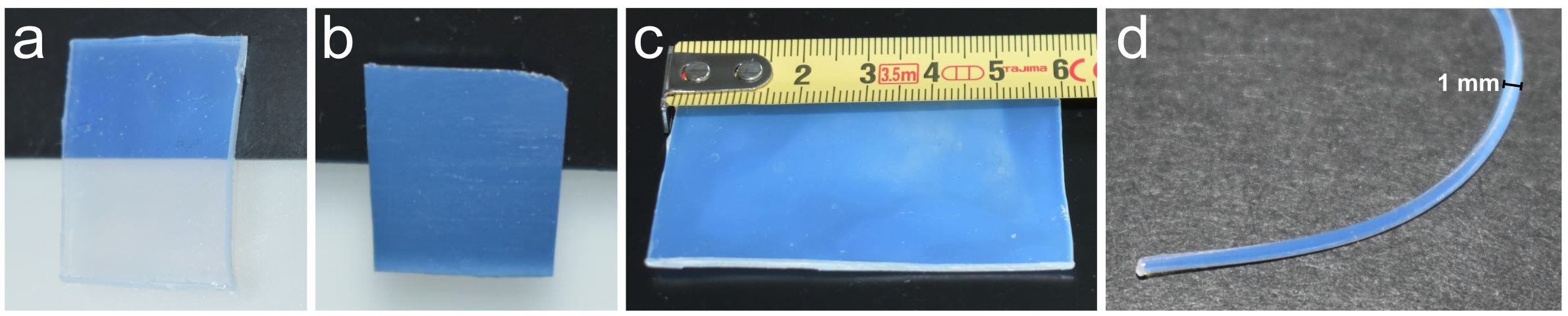}
\caption{\emph{Color in samples of different size and geometry.} a) A blue phase-separated sample formed from transparent PS (Sample 4, $X=0.20$) across a white and a black background. Left vertical side of the sample: 1.5 cm. b) A blue phase-separated sample formed from black PS across a white and a black background. The color exhibits no background dependence. Left vertical side of the sample: 1 cm. c) A blue phase-separated sample about $\sim 20~\mathrm{cm^2}$.  c) A blue phase-separated fiber. Diameter: 1 mm.}
\label{fig:samples}
\end{figure*}

\par Our PS-PMMA samples are translucent.  
Therefore, their color is more evident on a dark background than on a light background,  as shown clearly in Figure \ref{fig:samples}a.  
We can mitigate this effect by incorporating a low concentration of broadband absorber into the  composite \cite{forster2010biomimetic}. 
Figure \ref{fig:samples}b shows a sample where the same phase separation process was executed in a commercial black polystrene sample.
With this built-in absorption, the dependence of the appearance on the background disappears.

\par Structural colors based on thin films or coatings with specific thickness can be challenging to apply to curved surfaces and large areas.  Our approach to polymerization induced phase-separation is easy to apply to a wide variety of sample geometries. 
As a proof of concept, we show that our process can create larger structurally colored sheets (Figure \ref{fig:samples}c) and fibers (Figure \ref{fig:samples}d).

\par 
In conclusion, we show that it is possible to produce a structurally colored material based on solid-state phase separation driven by free-radical polymerization. 
While the resulting structures are stable and have the appropriate length scale, they are too disordered to reflect  a narrow range of wavelengths.  
The palette of structural colors is currently limited to blue and white, nevertheless this approach offers many potential benefits.  
First, in contrast to most self-assembly strategies, phase separation processes can generate well-defined supermolecular structures in a single step. 
Second, solid-state polymerization-induced phase separation is inherently self-limiting.  
As monomer is consumed, the matrix hardens and arrests coarsening. 
Third, photonic nanostructures can be easily incorporated into a wide variety of sample geometries.
Finally, this physical approach to arrest phase separation should extend to a wide range of chemistries.
We expect that this process can be extended to any pair of immiscible polymers, where one monomer is a good solvent for the second polymer.

\par In order to achieve a broader range of more saturated colors, as found in bird feathers, we need to achieve stronger structural correlations.
Free radical polymerization provides little control of the polymerization kinetics.
In solution, it is well known to create a very broad distribution of chain lengths. 
By contrast, living radical polymerization methods (such as atom transfer radical polymerization (ATRP) or reversible addition fragmentation chain transfer (RAFT)) provide tight control of the polymerization process and yield narrow molecular weight distributions \cite{wang1995controlled, matyjaszewski2018advanced}. 
By adapting these chemistries to the solid state, we expect that we can gain sufficient control of the nucleation and growth process to produce structures with more sharply defined length-scales, and tunable, saturated colors.

\section*{Experimental Section} 
\textit{Preparation of the transparent PS samples}. Polystyrene pellets, average Mw $\sim 280,000$ Da (CAS: 9003-53-6) were purchased from Sigma-Aldrich. 
They were dried in vacuum at \SI{60}{\celsius} overnight in a Binder Vacuum drying oven, and either hot-pressed into films of thickness 1 mm using a Fontune Holland table press, or extruded into fibers with a Dr. Collin GMBH extruder. 
Pressing parameters: \SI{205}{\celsius} for 5 min, 40 kN. Extrusion parameters: \SI{200}{\celsius}, 3 rpm.
The films or fibers were then cut into pieces and annealed in vacuum at \SI{103}{\celsius} in the Binder oven to release stresses. \\
The black polystyrene (1 mm thick) used in Figure \ref{fig:samples}b was used as purchased from Kitronik Ltd. The material consists of high impact polystyrene (HIPS) covered with a layer of general purpose polystyrene (GPPS). Product code: 5701. Phase separation happened in the GPPS layer. \\
\textit{Fabrication of the structurally colored composites}. 
Methyl methacrylate 99\% from Sigma-Aldrich (CAS: 80-62-6) containing monomethyl ether hydroquinone as inhibitor was purified through filtration in a chromatographic column containing an inhibitor remover from Sigma-Aldrich (Product code: 311332). The monomer was mixed with ethanol 96\% (Sigma-Aldrich, CAS:64-17-5) and a thermal free radical initiator, 2,2'-Azobis(2-methylpropionitrile) >98\% (Sigma-Aldrich, CAS:78-67-1). The ratio between monomer and initiator was fixed to 0.25 g AIBN in 10 ml MMA. The volume concentration of MMA in ethanol varied between $35\%$ and $50\%$. Transparent polystyrene films or fibers were then soaked into the solutions and left to equilibrate for at least 24 h in a container with a sealed cap. 
The container was then put in the oven (VWR, Venti-Line) at the desired $T_p$ for 90 min.
After the phase separation occurred, the containers were removed from the oven and the samples extracted from the solution, dried with a towel and left on a glass substrate at room temperature over night. \\
\textit{Reflectivity measurements}. The samples were mounted on a stage and illuminated using an optical fiber with collimated white light from a DH-2000-BAL Deuterium-Halogen Lightsource from Ocean Optics. The incident light was normal to the sample surface. The size of the illuminated spot could be tuned using an aperture placed between the collimating lense at the surface of the optical fiber and the sample, and it was set to $\sim 3$ mm in diameter. A detector was positioned at \SI{15}{\degree} from the direction of illumination. The detector consisted of an aperture, a collecting lens and a second optical fiber connected to a QE-Pro High Performance Spectrometer from Ocean Optics. At the start of the measurement, a reference spectrum of the incident light was collected with no sample in place. A spectrum with the \NEW{shutter closed to block the incident} light (dark noise) was also acquired. The reflectivity was calculated as the ratio between the reflected intensity minus dark noise and the reference spectrum minus dark noise. The plot of intensity as a function of wavelength was smoothed with a moving average filter. \\
\textit{NMR analysis}. Sample material was dissolved in chloroform-d >99.8\% from Apollo Scientific (CAS: 865-49-6). $^{1}$H-NMR measurements were performed using a Bruker UltraShield 300 MHz magnet, and analyzed using the software MestReNova. For the compositional analysis at different locations in the same sample, the surface layer was removed from the sample using a razor blade, and the remaining material along the sample thickness was considered to be the bulk. \\
\textit{STEM imaging}. Thin sections of 60 nm were obtained from the samples with a diamond knife (Diatome Ltd., Switzerland) on a Leica UC6 ultramicrotome (Leica Microsystems, Heerbrugg, Switzerland), and placed on Formvar and carbon coated TEM grids (Quantifoil, Großlöbichau, Germany). The sections were then coated with 5 nm of carbon using a Carbon Evaporator Safematic CCU-010. STEM analysis was performed using a ThermoFisher (FEI) Magellan 400 electron microscope. Electron micrographs of bird feathers were acquired according to reference \cite{prum2009development}. \\
\textit{SAXS experiments}. SAXS experiments on polymers were performed at the cSAXS (X12SA) beamline at the Swiss Light Source (SLS, Paul Scherrer Institut). Beam energy: 13.6 keV, i.e. wavelength $\lambda = 0.91$ Å. The intensities were recorded with a Pilatus 2M detector (Paul Scherrer Institut). Sample-detector distance: 7.1 m. The wave number ($q$) range was $0.01$ – $1~\mathrm{nm^{-1}}$. The beam diameter was approximately $300~\mathrm{\mu m}$. 
The q range was calibrated using silver behenate. The samples were taped to a metallic grid, so that the area probed by x-rays was free standing. The measurements occurred at several positions on the samples using 1 s exposure, and the median was then calculated. SAXS experiments on bird feathers in transmission geometry were collected at beamline 8-ID at the Advanced Photon Source as described in reference \cite{saranathan2012structure}.
\NEW{\textit{Matlab analysis of STEM images.} First, the PMMA inclusions had to be detected. A band-pass filter was applied. The high value of the filter was chosen as the highest value at which the number of detected particles deviates from a linear trend. An intensity threshold is then applied. The value of this threshold is determined as 95\% of the mean intensity across all the detected particles. Second, once the particles are defined, the determination of centroids and the calculation of the size distribution are performed using standard Matlab functions, such as regionprops. 
} \\

\section*{Conflicts of interest}
There are no conflicts of interest to declare.

\section*{Author contributions}
Each author contributed to this work as follows: E.R.D. and  R.M.R. initiated the project. A.S., E.R.D., G.P., M.F., V.S. designed polymer experiments.  A.S., A.M., D.M. and R.G. performed polymer experiments.
A.S., E.R.D., A.M. and D.M. analyzed polymer data. V.S. and R.P. designed, performed and analyzed all the bird feather experiments. A.S. and E.R.D. wrote the manuscript with input from all authors.

\section*{Acknowledgements}
We thank Hui Cao for helpful conversations, as well as Maja Margrit Günthert and Joakim Reuteler (ScopeM, ETH Zurich) for their help with sample preparation for STEM,
Thomas Schweizer and Christian Furrer for their help with fiber extrusion, Nicolas Bain and Dominic Gerber for their help with programming, \NEW{Tianqi Sai and Jinyoung Kim for their support during the SAXS experiment}, and 
Kirill Feldman for his inspirational skepticism and support with materials processing in
the Laboratory of Soft Materials (Department of Materials, ETH Zurich).
The authors acknowledge the Paul Scherrer
Institut, Villigen, Switzerland for provision of synchrotron radiation beamtime at the cSAXS beamline of the SLS. 
SAXS data on bird feathers were collected with the help of Alec Sandy and Suresh Narayanan at beam line 8-ID-I of the Advanced Photon Source at Argonne National Labs, and supported by the US Department of Energy, Office of Science, Office of Basic Energy Sciences, under Contract No. DE-AC02-06CH11357.

\newpage

\vspace{2cm}

\section*{Supporting Information}
\renewcommand{\thefigure}{S\arabic{figure}}
\setcounter{figure}{0}

\begin{figure*}[h!tb]
\centering
\includegraphics[width=8cm]{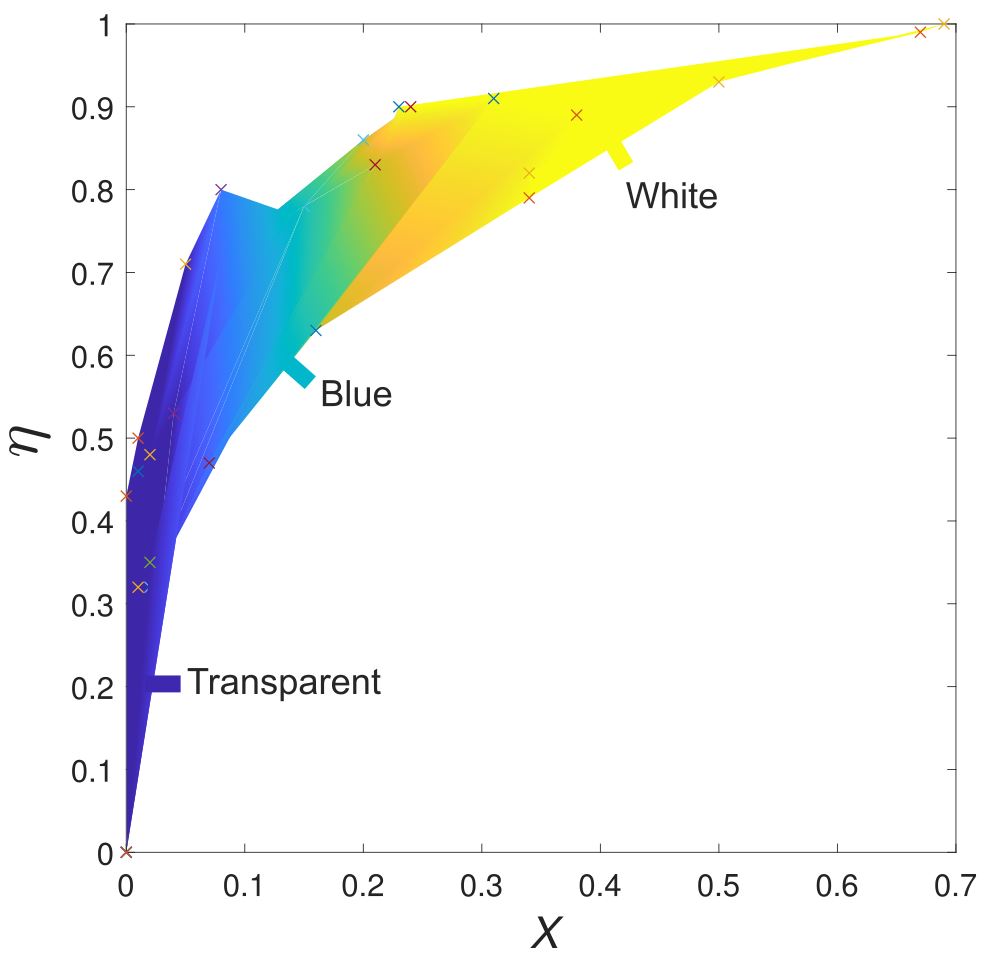}
\caption{\emph{Sample color as a function of PMMA fraction ($X$) and monomer conversion ($\eta$).} The colored area indicates the explored compositional space. Each 'x' indicates the composition of a sample. Regions corresponding to different colors are labelled.
}
\label{fig:SInmr}
\end{figure*}

\begin{figure*}[bt]
\centering
\includegraphics[width=11cm]{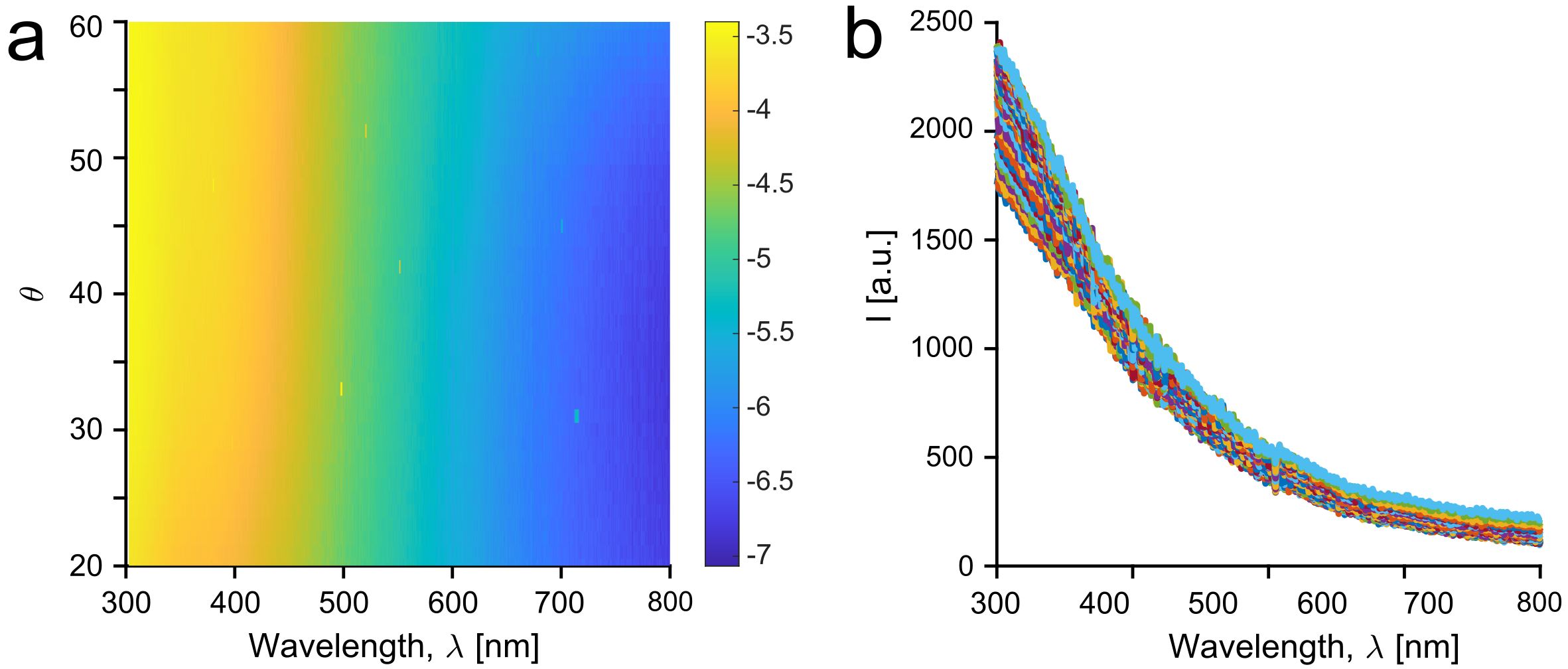}
\caption{\emph{Angle-independent color.} The color of the PS-PMMA composites is angle-independent. a) Logarithm of the intensity as a function of wavelength and detection angle, $\theta$. As $\theta$ increases, the intensity of the spectra at specific wavelengths decreases slightly. On the right: color bar. b) The same spectra as a function of wavelength. The measured intensity is reported on the vertical axis. Each line corresponds to a different detection angle $\theta$.
}
\label{fig:SIangle}
\end{figure*}

\begin{figure*}[bt]
\centering
\includegraphics[width=15cm]{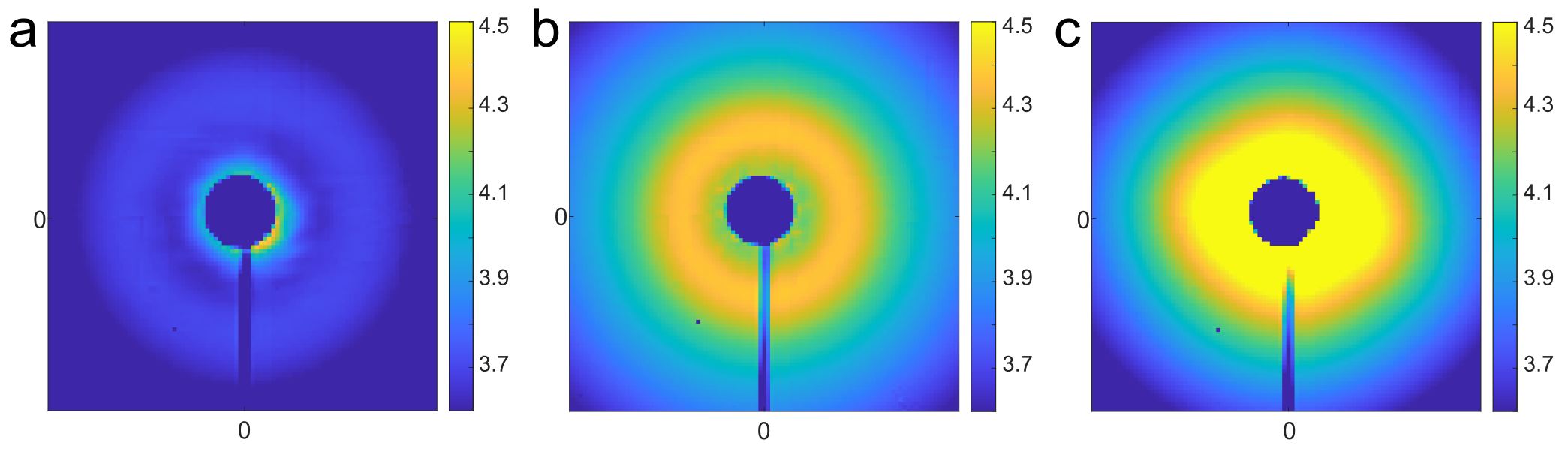}
\caption{\emph{SAXS patterns}. Complete small-angle X-ray scattering patterns for samples a) \#1, b) \#2, and c) \#3. The dark feature in the middle is the beamstop. Each axis ranges from -0.05 to $+0.05~\mathrm{nm^{-1}}$. On the right of each image: color bar. 
}
\label{fig:SIsaxs}
\end{figure*}

\begin{figure*}[bt]
\centering
\includegraphics[width=6cm]{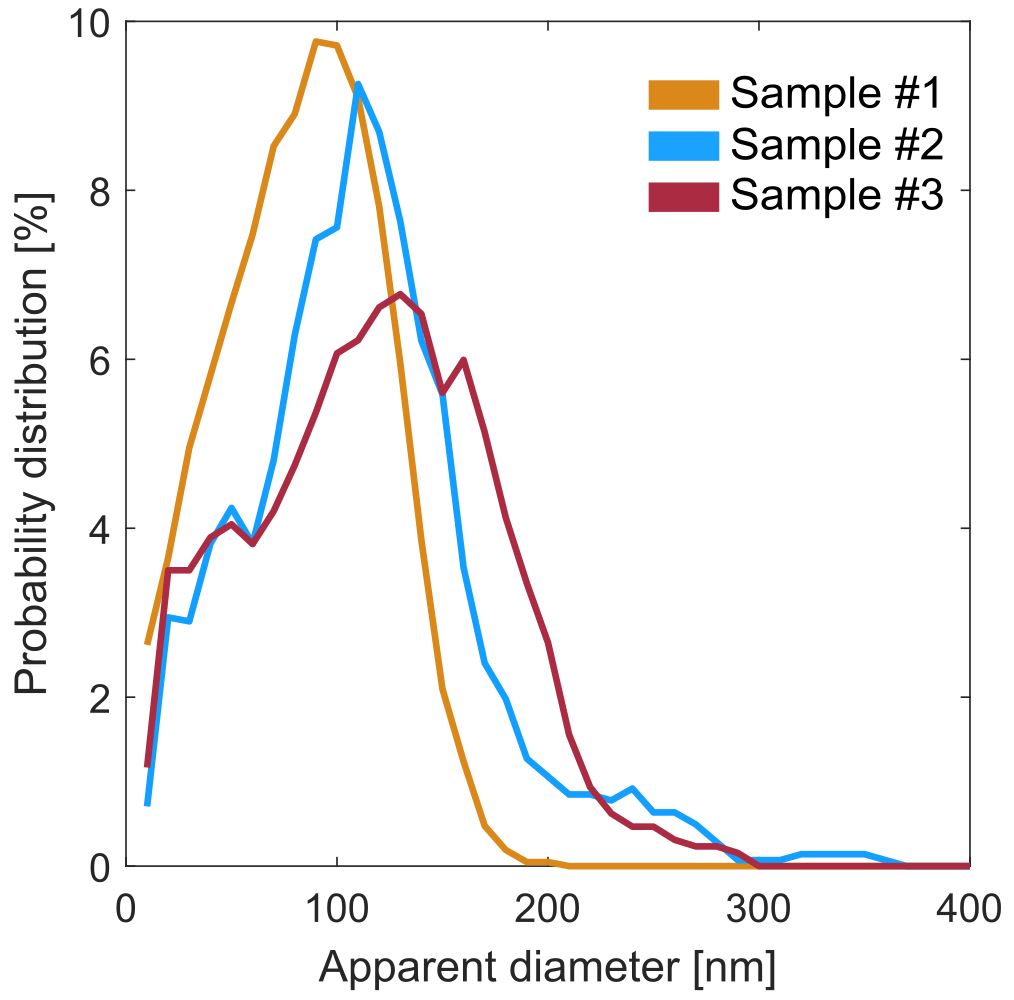}
\caption{\emph{Size distribution of PMMA inclusion}. Probability distributions of the diameters of PMMA inclusions calculated from the STEM images in Figure 4a for samples \#1, \#2 and \#3. "Apparent" because the images section a 3D material along a 2D plane, and this affects the perceived size distribution of the spherical inclusions.  
}
\label{fig:SIsize}
\end{figure*}



\end{document}